\documentclass[twocolumn,aps,pre,showpacs,preprintnumbers,amsmath,amssymb]{revtex4}
\usepackage{epsf}

\begin{document}
\draft
\newcommand{\ve}[1]{\boldsymbol{#1}}

\title{CO-activator model for reconstructing Pt(100) surfaces: local microstructures and chemical turbulence}
\author{Natalia Pavlenko}
\address{Institute for Condensed Matter Physics, Svientsitsky str. 1, 79011 Lviv, Ukraine}
\date{\today}

\begin{abstract}
We present the results of the modelling of CO adsorption and catalytic CO oxidation on inhomogeneous 
Pt(100) surfaces which contain structurally different areas. These areas are formed during the CO-induced 
transition from a reconstructed phase with hexagonal geometry 
of the overlayer to a bulk-like ($1 \times 1$) phase with square atomic arrangement.
In the present approach, the surface transition is explained in terms of nonequilibrium bistable behavior. 
The bistable region is characterized by a coexistence of the hexagonal and ($1 \times 1$) phases and
is terminated in a critical bifurcation point which is located at ($T_c \approx 680$~K, $p_{\rm CO}^c \approx 10$~Torr).
Due to increasing fluctuations, the behavior at high temperatures and pressures in the 
vicinity of this cusp point should be qualitatively
different from the hysteresis-type behavior which is typically observed in the experiments
under ultrahigh vacuum conditions.
On the inhomogeneous surface, we find a regime of nonuniform oscillations 
characterized by random standing waves of adsorbate concentrations. 
The resulting spatial deformations of wave fronts 
allow to gain deeper insight into the nature of irregular oscillations on Pt(100) surface.

\end{abstract}


\maketitle

\section{INTRODUCTION}

The processes of surface reconstruction play a central role in heterogeneous catalysis \cite{white}.
A prominent example of a catalyst with the reactivity strongly affected by the
rearrangement of the substrate atoms is the Pt(100) surface. The 
mechanism of such rearrangements is the tensile excess stress due to charge depletion of 5d orbitals
of surface Pt atoms \cite{fiorentini}. As a result, the bulk-like ($1\times 1$) termination
of clean Pt(100) corresponding to the square configuration of Pt atoms becomes metastable 
which causes a transformation to the hexagonal (hex) surface atomic arrangement. 
From the point of view of the catalytic properties, the crucial importance has the fact that
such surface reconstructions are strongly affected by the adsorbates. The CO-induced lifting of the hex surface 
reconstruction and the stabilization of the $(1 \times 1)$ phase
is forced by a gain in the CO adsorption energy and proceeds through the
nucleation of ($1 \times 1$)-islands with high CO coverage \cite{behm,thiel,hopkinson}.

The reconstruction of Pt surfaces is a key factor responsible for their complex oscillating behavior and pattern
formation \cite{imbihl}. This is also supported by the fact that the stable Pt(111) surface where the reconstruction 
is not observed, does not exhibit oscillation properties. 
During the reconstruction, the surface of Pt(100) becomes inhomogeneous and
contains coexisting micrometer-size regions of different (hex and square) atomic configurations. 
As demonstrated in low energy electron diffraction (LEED) studies \cite{cox,eiswirth,imbihl2}, the 
propagating waves of surface modification can result in highly irregular 
spatio-temporal character of the oscillations. 

Even with a constant oscillation period, the time variation of the local LEED intensity of small surface spots
differs from the temporal behavior of the intensities integrated over the whole surface area \cite{eiswirth}.
Such a behavior is in contrast to the uniform oscillations on Pt(110) surfaces 
originating from the global coupling of various surface patches through the gas phase \cite{imbihl,eiswirth}.  
It should be noted that the mechanism of the global coupling has been extensively studied in a series of 
experimental works and by the theoretical modelling \cite{falcke,bar2,mertens,veser,khrustova,thostrup,monine}. 
In distinction to Pt(110), the irregular oscillations during CO oxidation on Pt(100) possibly originate from
the propagating chemical reaction fronts which can be triggered by defects and other surface
imperfections \cite{cox,imbihl4}.
The kinetic Monte Carlo simulations performed for the heterogeneous Pt(100) surfaces containing 
mesoscopic hex regions in macroscopic
($1 \times 1$) surrounding, have essentially shown local unsynchronized oscillations\cite{kortluke}. These local 
oscillations disappear on the $\mu$m--scale, whereas the experiment demonstrates well developed irregular oscillations even 
on the macroscopic mm-scale \cite{eiswirth}. Another aspect which can lead to the irregular temporal character of oscillations 
is the stong fluctuations on nm-size catalyst particles with (100)-facets \cite{zhdanov}. As the fluctuations in surface 
coverages are increasing with the decrease of the system size \cite{pavlenko,zhdanov2}, they cannot satisfactorily 
explain the irregularities observed on single crystal surfaces of Pt(100).  

To gain deeper insight into the nature of the irregular behavior, 
in the present work we consider a modelling of 
spatially inhomogeneous Pt(100) surfaces. In our approach, we account for the fact 
that the hex-square transformation 
occurs through a hysteresis and has a character of a first-order phase transition. 
Under the non-equilibrium conditions,
the hysteresis features are described in terms of a bistable behavior
which is characterized by a coexistence of the hex and square phases
in a wide range of temperatures and partial pressures. Such a mechanism 
allows to obtain in a natural way  
the inhomogeneous surface state, without introducing additional separate variables 
for the description of the adsorbate coverage in different phases. 
Upon the increase of temperature and CO pressure, the bistable region narrows 
and terminates at a bifurcation cusp point ($T_c\approx 680$~K, $p_{\rm CO}^c\approx 10$~Torr). Due
to strong fluctuations near this critical point, the behavior at 
high temperatures and pressures should be qualitatively different from the hysteresis-type behavior 
observed on Pt(100) under ultahigh vacuum conditions\cite{thiel}. 

Furthermore, we address two main questions related to
the surface inhomogeneities. First, we analyze the role
of the inhomogeneities in the nucleation and growth of adsorbate islands.
We find that the difference between the CO desorption rates for the hex and square 
substrate geometries is responsible for the trapping of the adsorbed CO \cite{behm}, 
a property responsible for the growth of CO islands 
during the hex-square transformation.
Second, we study the oscillation behavior on inhomogeneous surface 
during the reaction of CO oxidation. 
We show that the surface inhomogeneities lead to the Benjamin-Feir instability
and to a new inhomogeneous oscillating regime. This new regime is characterized by the random 
standing waves in spatially separated surface regions. We investigate the driving mechanisms
of irregular oscillations and provide the analysis of the stability of the system with respect to nonuniform 
perturbations. 
The obtained deformations of the chemical wave fronts appear to be a direct consequence of structural 
inhomogeneities typical for Pt(100), and can be a possible clue for the explanation 
of the irregular oscillating behavior on this type of surfaces.

\section{CO-ACTIVATOR MODEL AND ITS ANALYSIS} \label{act_model}

A fundamental property related to the adsorption of CO on low-index Pt(100)
surfaces is a transformation of the initially stable reconstructed
hex Pt surface into the bulk-like ($1 \times 1$) structure.
The lifting of the surface reconstruction is activated in the case when the concentration
(coverage) of adsorbed CO exceeds a critical value $\Theta_{\rm CO}^1 \approx 0.05$~ML.
In a wide range of the CO surface coverages $\Theta_{\rm CO}^1 < \Theta_{\rm CO} <
\Theta_{\rm CO}^2$ ($\Theta_{\rm CO}^2 \approx 0.3$~ML), the surface 
contains the coexisting areas of the ($1 \times 1$)- and hex-geometry \cite{hopkinson}.
In the LEED experiments, the range of temperatures and CO partial pressures 
determining the phase coexistence is identified by a hysteresis behavior of LEED
intensities \cite{thiel}.

It is noteworthy that in the nonequilibrium conditions, the hex-($1 \times 1$) coexistence
should be described in terms of a bistable behavior.
To provide a consistent analysis of the CO controlled bistability,
we introduce a variable $A=\Theta_{\rm CO}$ for the description of
the CO surface coverage and a parameter $n$ which measures a degree of the surface
reconstruction. The value $n=0$ corresponds to a fully reconstructed
hex surface whereas the maximal value $n=1$ describes the recovered homogeneous 
($1 \times 1$) state.

Within the two-variables approach, the process of the CO-induced surface
transformation contains two main stages.
The first stage includes the change of the local mesoscopically averaged CO surface
coverage during the steps of CO adsorption, desorption 
and CO diffusion on Pt surface:
\begin{eqnarray} \label{a}
\frac{\partial A}{\partial t}=p_A k_A (1-\left({A}/{A_s}\right)^q) s_A -d_A\cdot A + D_A
\Delta A.
\end{eqnarray}
Here $p_A$ and $k_A$ denote the partial pressure and the impingement rate
of CO, respectively, and $A_s=0.5$ refers to the
maximal (corresponding to a saturation) CO coverage; $D_A$ is the CO diffusion coefficient.
The factor $q>1$ models the precursor-type kinetics of the CO adsorption \cite{krischer,gasser}. 
In our analysis, a typical value $q=3$ for the precursor factor is chosen.

For the kinetic processes on the inhomogeneous surface containing a mixture of
hex- and ($1 \times 1$)-islands, we need to consider two different contributions to 
the adsorption/desorption steps in (\ref{a}). 
These two different types originate from the processes CO$_{\rm gas}\leftrightarrow$CO$_{\rm hex}$ and 
CO$_{\rm gas}\leftrightarrow$CO$_{\rm 1 \times 1}$ 
which contribute with the weights $(1-n)$ and $n$ to the total rates
of surface adsorption and desorption. 
As a consequence, the final expression for the CO sticking and CO desorption
coefficients $s_A$ and $d_A$ contain both contributions 
\begin{eqnarray} \label{sa}
&& s_A=s_A^0 (1-n)+s_A^1 n,\\
&& d_A=d_A^0 (1-n)+d_A^1 n,\nonumber
\end{eqnarray}
where $d_A^0=\nu_0 \exp(-E_d^0/RT)$ and $d_A^1=\nu_1 \exp(-E_d^1/RT)$.
In the expressions for $s_A$ and $d_A$, the parameters 
$s_A^0$, $d_A^0$ and $s_A^1$, $d_A^1$ are the CO sticking and desorption 
coefficients in the hex and ($1\times 1$) phases, respectively. 
As is shown in \cite{hopkinson}, the low desorption rate $d_A^1$
is a prime factor responsible for high 
CO coverages of the ($1\times 1$) surface.
In contrast, due to higher desorption rate on the hex surface, 
significant part of adsorbed CO molecules 
desorps which leads to substantially lower CO coverages in the hex state.   

For the sticking coefficients, we use the values
$s_A^0=0.75$ ($A=0.05$, hex geometry) and $s_A^1=0.34-0.7$ ($A=0.5-0.3$, $1 \times 1$ geometry)
reported in Ref.~\onlinecite{thiel}. In the Arrhenius-type form of the 
desorption coefficients $d_A^0$ and $d_A^1$ we choose the following values for the rate parameters: 
$\nu_0=4\times 10^{13}$~s$^{-1}$ and $\nu_1=3\times 10^{15}$~s$^{-1}$; 
$E_d^0=27.5$~kcal/mol and $E_d^0=34.5$~kcal/mol \cite{thiel}. 

Furthermore, in our approach we consider the transformation of the
substrate state activated by the adsorbed CO. In the two-variable model, the time evolution of 
the local mesoscopically averaged surface
state parameter $n$ is described by the following phenomenological equation:
\begin{eqnarray} \label{n}
&&\frac{\partial n}{\partial t}= \gamma_{01} \cdot n(1-n) (A/A_s) \cdot \Theta(A-A_c^1) \nonumber\\
&& -\gamma_{10} \cdot n (1-A/A_s)^2 + \kappa \Delta n.
\end{eqnarray}
Here the first term refers to the lifting of the
hex reconstruction. The rate of the lifting is proportional to the CO coverage $a$ 
and to the factor $\gamma_{01}n(1-n)$
which is a mean-field form of the effective flux between the hex and ($1\times 1$)- regions.
By introducing the step function $\Theta(A-A_c^1)$ ($\Theta(\xi)=1$ for $\xi >0$
and 0 otherwise), we account for the experimental observation that the transition to
the ($1 \times 1$) phase does not occur below a critical coverage
$A_c^1=\Theta_{\rm CO}^1$. In our calculations, we describe the transition rate
$\gamma_{01}=\nu_{01}\exp (-E_{01}/RT)$ by an Arrhenius-type expression where the activation
energy $E_{01}\sim 2$~kcal/mol corresponds to a small activation barrier
for the transformation from the hex- to the ($1 \times 1$)-state \cite{thiel},
and the prefactor $\nu_{01} \approx 10$~s$^{-1}$ 
gives correct time scales for the transformation in the considered temperature
interval \cite{krischer,bar}.

The second term in (\ref{n}) refers to the reverse ($1\times 1$)$\rightarrow$hex
reconstruction with a rate coefficient $\gamma_{10}$. Here the
factor $(1-A/A_s)^2$ accounts for the property that the
reconstruction can occur locally if a small cluster of neighbouring surface positions
(a pair in the simplest case)
is not occupied by adsorbed CO which is in analogy with the assumption
considered in \cite{kortluke,kuzovkov}. 

For an inhomogeneous surface, the laplacian term $\Delta n$ in (\ref{n}) originates from the
contribution of the interfaces between different surface geometries to the total 
system energy \cite{bray,hildebrand}.
Consequently, the coefficient $\kappa$ describes the energy costs of such 
interfaces and is related 
to the characteristic interface width $L_n=\sqrt{\kappa/\gamma_{01}}$ which is typically of the
order of few nanometers. 
We note that the characteristic diffusion length
of the adsorbed CO is given by $L_A=\sqrt{D_A/\gamma_{01}}$ and is of the order of several micrometers.
The fast CO diffusion between the hex- and ($1\times 1$)-areas described by the
diffusion term in (\ref{a}), controls the $\mu$m size 
of CO islands which is discussed in details in section~\ref{nucleation}.   

The equations (\ref{a}) and (\ref{n}) can be transformed by the substitution:
\begin{eqnarray} \label{scale}
&& A=a\cdot A_s;\, x=\tilde{x}L_A;\, \tilde{d}_A=d_A/\gamma_{01};\,
\tilde{t}=t\gamma_{01};\nonumber\\
&& \tilde{p}_A=p_Ak_A/A_s\gamma_{01}; \, \eta=\kappa/D_A; \,
\gamma=\gamma_{10}/\gamma_{01}
\end{eqnarray}
into the following dimensionless form
\begin{eqnarray} \label{an}
&& \frac{\partial a}{\partial \tilde{t}}=f_1(a,n)+\Delta_{\tilde{x}}a\\
&& \frac{\partial n}{\partial
\tilde{t}}=f_2(a,n)+\eta\Delta_{\tilde{x}}n,\nonumber
\end{eqnarray}
where
\begin{eqnarray} \label{f12}
&& f_1=\tilde{p}_A (1-a) s_A-\tilde{d}_A \cdot a,\\
&& f_2=n(1-n)a\cdot \Theta(a-a_c^1)-\gamma n (1-a)^2. \nonumber
\end{eqnarray}
For a given
$a\ne 0$, the second equation (\ref{an}) leads to the two stable steady
states: $n_1=0$ and $n_2= 1-\gamma (1-a)^2/a$. In the limit
$a \rightarrow 1$, the second solution approaches the value $n_2=1$
which implies the full lifting of the hex geometry by the CO adsorption. In the following 
discussion, we will omit for convenience the tilde signs in the notations of 
dimensionless time and coordinate.  

A remarkable property of the system (\ref{sa}), (\ref{an}), (\ref{f12}) is a strong
dependence of the CO adsorption and desorption rates on the local configurational state of
the surface determined by the parameter $n$. 
In analogy to \cite{hopkinson}, this dependence can can be conveniently analyzed
in terms of a net sticking probability of CO molecules defined as
\begin{equation} \label{nets}
s_A^n=s_A-\tilde{d}_A a /\tilde{p}_A (1-a).
\end{equation}
Specifically, despite a large initial sticking coefficient $s_A^0$ in the 
hex regions, the significant contribution of the second desorption term in (\ref{nets})
results in a suppression of $s_A^n$   
which reflects the fact of low CO coverages $A \le 0.05$~ML observed 
on the reconstructed surfaces.
In contrast, due to the low desorption rates on the ($1\times 1$) substrate,  
the adsorption of CO will result in large values of $s_A^n$ and consequently in 
high CO coverages in the regions with the lifted 
reconstruction.
In the following analysis, we will use the net sticking probability as a 
relevant quantity for the interpretation
of the growth kinetics of CO islands on inhomogeneous substrates.

\subsection{The surface transition hex-($1 \times 1$) on Pt(100): bistable behavior}

A central feature of the model (\ref{an}),
(\ref{f12}) is the occurrence of a bistable region characterized by 
the coexistence of the states $n_1$ and $n_2$ with low and high CO coverage. 
In the analysis of (\ref{an}) and (\ref{f12}), we choose
$k_A=1.13\cdot 10^{5}$~s$^{-1}$Torr$^{-1}$ML and
perform the calculations of the steady states using the rate parameter
$\gamma_{10}=\nu_{10}\exp{(-E_{10}/RT)}$ with $E_{10}=23$~kcal/mol and $\nu_{10}=10^{12}$~s$^{-1}$.
The large value of $E_{10}$ is consistent with the high activation barrier
for the hex reconstruction of the clean Pt(100) surface reported in \cite{thiel}.  
In the Arrhenius form of the desorption rate parameter $d_A^1$ of the ($1\times 1$) substrate, we choose
the values $\nu_1=1\times 10^{15}$~s$^{-1}$ and $E_d^{1}=37$~kcal/mol. These values
have been deduced in \cite{thiel} from the temperature dependences of 
the corresponding LEED intensities which exhibit a hysteresis-type behavior. 

\begin{figure}[ht]
\epsfxsize=8.0cm {\epsffile{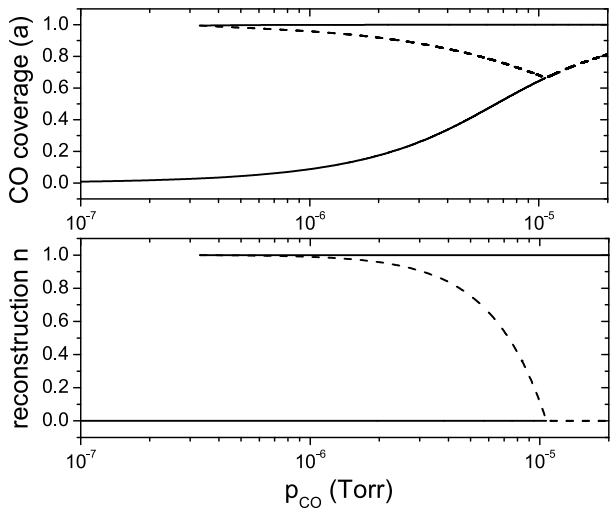}} \caption{Stationary CO coverage $a$ and
surface reconstruction parameter $n$ versus CO pressure. 
Here $T=450$~K. The unstable solutions correspond to
the saddle nodes and indicated by dashed curves.} \label{fig1}
\end{figure}

Fig.~\ref{fig1} shows the stationary solutions calculated for different CO pressures.
Here the bistable region where the stable hex and ($1\times 1$) phases coexist is
located in the range of pressures between $p_A^1=3\cdot 10^{-7}$~Torr and
$p_A^2=10^{-5}$~Torr.

\begin{figure}[ht]
\epsfxsize=8.0cm {\epsffile{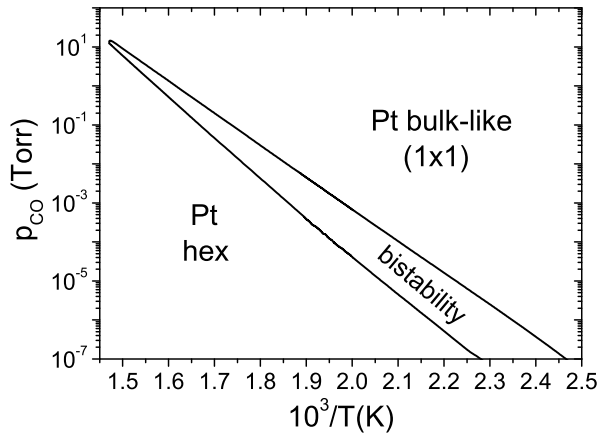}} \caption{Phase diagram ($1/T$, $p_{\rm
CO}$) showing the regions of the stability of the hex and ($1 \times 1$) phases 
with the indication of the bistable
region where both these phases coexist on Pt surface. 
} \label{fig2}
\end{figure}

The regions of the stability of different phases together with the bistable region
are indicated on the phase diagram ($1/T$, $p_{\rm CO}$) shown in Fig.~\ref{fig2}. We note that the
topology of the diagram and the location of the bistable region agree well
with a LEED diagram obtained by the estimation of isosteric heats of CO adsorption \cite{thiel}. 
A remarbable feature of our diagram is that the bistable region narrows with the increasing $p_{\rm CO}$ and
terminates at a critical bifurcation point at $p_{\rm CO}^c \approx 2\times 10^{1}$~Torr and $T_c\approx 680$~K. 
The existence of such a cusp point on the diagram is a clear manifestation of 
the critical behavior which is associated with a strong increase of 
near-critical fluctuations in the system. We should emphasize that in this high pressure range,
the direct comparison with the experimental data obtained at low CO pressures up to $10^{-4}$~Torr 
cannot be provided. In view of this, additional experimental studies conducted 
at high pressures and temperatures are required in order to
verify the existence of the obtained critical bifurcation point.
For Pt(100), such experiments would be of a central importance
in bridging a "pressure gap" between the hysteresis-type behavior at low $p_{CO}$
and possible critical fluctuations which should exist at high CO pressures. 

\section{ISLANDS NUCLEATION ON INHOMOGENEOUS Pt SURFACE} \label{nucleation}

In the bistable region, the problem of prime interest is the
behavior of adsorbate molecules on a surface which contains micrometer-size areas of
different geometry. 
To study the nucleation and growth of adsorbate islands on such an 
inhomogeneous substrate, we have to
consider the full set of the reaction-diffusion equations
(\ref{an}) and (\ref{f12}). It should be noted that the strong difference between 
the characteristic nanoscale
width of the hex/($1 \times 1$) interfaces $L_n$ and the micrometer scale of the
CO diffusion length $L_A$ results in small values of the ratio $\eta$ in (\ref{scale}) which 
is of the order $10^{-3}$--$10^{-4}$. Furthermore, in our analysis 
we choose a typical value $\eta=0.005$. 
The equations (\ref{an}) and (\ref{f12}) have been solved numerically
using an implicit two-level finite-difference scheme on a discrete finite grid 
with the Neumann boundary conditions
$\partial a/\partial x=0$, $\partial n/\partial x=0$. In this absolutely stable scheme, 
the standard finite difference approximations of the second-order accuracy 
for coordinate derivatives and the first order accuracy approximation 
for time derivatives are applied.

\begin{figure}[ht]
\epsfxsize=8.0cm {\epsffile{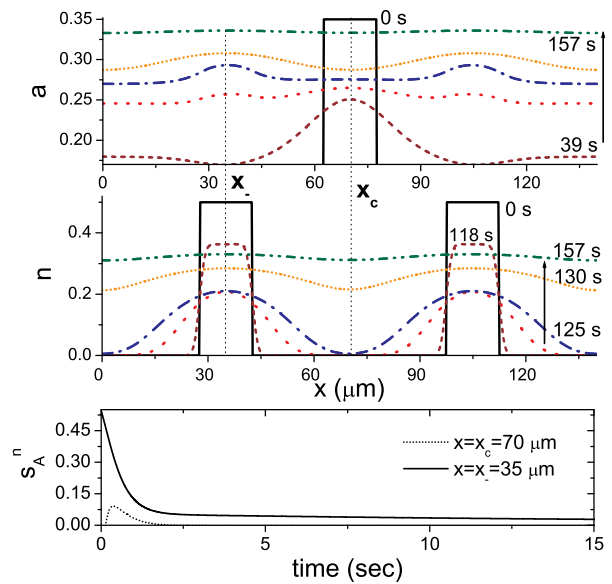}} \caption{Time evolution of spatial
profiles of $a$ and $n$ on inhomogeneous Pt surface for $p_A=3.3\cdot
10^{-6}$~Torr, $\eta=0.005$, $\gamma=0.5$ and $T=450$~K. The nucleation and growth of
two islands with high $a$-coverage occurs in the $1 \times 1$ surface
regions. The bottom panel shows the behavior
of the net CO sticking probabilities in the central hex ($x=x_c$) and in 
the unreconstructed ($x=x_-$) parts of the substrate.} \label{fig3}
\end{figure}

To demonstrate the influence of the surface geometry on the growth of adsorbate
islands, in Fig.~\ref{fig3} we present the spatio-temporal evolution of islands
with high CO coverage on inhomogeneous surface. For simplicity, we consider
a quasi-one-dimensional case described by the spatial coordinate $x$ which corresponds to
a stripe-like surface patterning.
The initial geometry of the substrate is characterized by two symmetric
regions exhibiting a high degree of transformation into the
($1 \times 1$) phase with $n=0.5$. These regions of the length 15~$\mu$m are located at a 
distance about 35~$\mu$m from the center and surrounded by
the reconstructed (hex)-areas with $n=0$. In the cental region, we initially
have a preformed CO island which is centered at $x=x_c$ and 
located in the hex-area beyond the ($1\times 1$)-patches. 
Furthermore, in the process of the temporal
evolution shown in Fig.~\ref{fig3}, the CO coverage of the central hex-area
decreases. This decrease is driven by the CO desorption which is dominant due to
the high desorption coefficient $d_A^0$. It is remarkable that the 
disappearance of the central CO island is accompanied by the simultaneous 
development of two new symmetric islands with high CO coverage. These two islands nucleate 
in the ($1\times 1$)-parts of the substrate where the desorption is suppressed
and CO molecules become trapped due to the high CO binding energy on the surface. 

The corresponding difference between the kinetic properties  
of different regions of inhomogeneous substrate can be interpreted in terms
of the time evolution of the net sticking probabilities $s_A^n$ shown in the bottom panel of
Fig.~\ref{fig3}. The predominant desorption of CO from the 
central hex area is reflected in the initial zero net sticking probability $s_A^n (x=x_c)$
which slightly increases with the expansion of the ($1\times 1$)-areas on the surface.
In contrast to this, in the symmetric regions with partially lifted surface reconstruction,
the net sticking probability is high due to low desorption rates on these parts of 
the surface ($s_A^n(x=x_-)$ in Fig.~\ref{fig3}). 

\begin{figure}[ht]
\epsfxsize=8.0cm {\epsffile{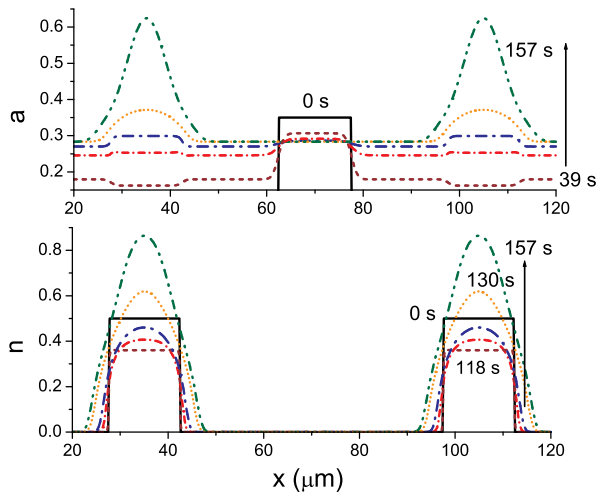}} \caption{Time evolution of spatial
profiles of $a$ and $n$ on inhomogeneous Pt surface for $p_A=3.3\cdot
10^{-6}$~Torr, $\gamma=0.5$ and $T=450$~K. The CO diffusion constant $D_A$ is by two orders
of magnitude smaller then that in Fig.~\ref{fig3}. To provide a correct
comparison of both spatial profiles, the parameter scaling is performed
relative to the reference data in Fig.~\ref{fig3}.} \label{fig4}
\end{figure}

In fact, the obtained simultaneous
disappearance and nucleation of CO islands in the areas of different geometry can be
characterized as an effective island propagation. The basic mechanism responsible
for this propagation, is the difference between the CO desorption rates
in the ($1\times 1$) and hex surface parts.
Consequently, in the ($1\times 1$) regions where the desorption rate is lower,
the rapid CO diffusion between the hex and ($1\times 1$) areas results in
the trapping of CO molecules and in the growth of the micrometer-size islands.

The effect of the CO diffusion on the island growth 
becomes more clear when we perform calculations with a small CO diffusion
coefficient which corresponds to the case shown in Fig.~\ref{fig4}. In
Fig.~\ref{fig4}, the decrease of $D_A$ by the two orders of magnitude corresponds
to the additional prefactor $10^{-2}$ near the diffusion term
in (\ref{an}). The comparison with Fig.~\ref{fig3} clearly shows that such a slow diffusion 
leads to a strong localization of the new symmetric CO islands 
in the ($1\times 1$)-areas and to a
significant slowing down of the hex-lifting process.

\begin{figure}[ht]
\epsfxsize=8.0cm {\epsffile{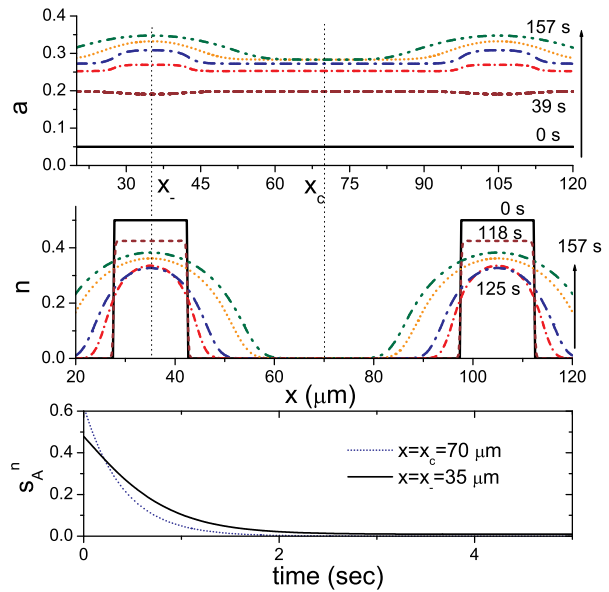}} \caption{Time evolution of spatial
profiles of $a$ and $n$ on inhomogeneous Pt surface for $p_A=3.3\cdot
10^{-6}$~Torr, $\gamma=0.5$ and $T=450$~K. The CO diffusion constant $D_A$ is by one order of
magnitude smaller then that in Fig.~\ref{fig3}. The initial spatial
distribution of CO is homogeneous with $a=0.05$. To provide a correct
comparison of both spatial profiles, the parameter scaling is performed
relative to the reference data in Fig.~\ref{fig3}. The bottom panel shows the behavior
of the net CO sticking probabilities in the central hex ($x=x_c$) and in
the unreconstructed ($x=x_-$) parts of the substrate.} \label{fig5}
\end{figure}

To analyze the role of the CO desorption in the island nucleation, 
we consider an inhomogeneous
surface where the initial CO coverage is uniformly distributed
with $a(x)=a_0=0.05$ (Fig.~\ref{fig5}). This is lower then
the critical coverage $a_c^1=0.1$ necessary for the hex$\rightarrow$($1\times 1$) transition. In the
course of temporal evolution, the CO coverage begins to increase predominantly in
the ($1\times 1$)-regions of substrate which leads to the development of two symmetric CO islands. 
Although the absolute sticking coefficient
in the ($1\times 1$)-areas $s_A^1$ is smaller that the hex sticking coefficient $s_A^0$, the low 
CO desorption rates lead to substantially higher net CO sticking probability 
in the ($1\times 1$)-regions, which is illustrated by the time evolution
of $s_A^n$ in the hex ($x=x_c$) and ($1\times 1$) ($x=x_-$) regions 
in Fig.~\ref{fig5}~(bottom panel). 
It should be noted that the obtained strong dependence of the net sticking probabilities
on the geometry of the substrate is in agreement with the scenario of CO island growth dynamics 
discussed in \cite{hopkinson}. In view of this, the difference in the CO desorption rates 
appears to be a central mechanism responsible for 
the CO island nucleation and CO trapping on inhomogeneous surfaces.

\section{MODELLING CATALYTIC CO OXIDATION}

The equations (\ref{a}) and (\ref{n}) represent in fact a CO-activator model
where the surface transition hex$\rightarrow$($1 \times 1$) is activated by the CO adsorption.
On the Pt(100) surfaces, this structural surface transformation is a key source for
complex spatial self-organization and temporal oscillations observed in
surface reactions like catalytic CO oxidation \cite{cox,eiswirth}. This is in
distinction to the oscillating behavior on Pt(110) surfaces where the oscillations
occur via a coupling through the gas phase. For Pt(100), the basic oscillation
mechanism consists in a periodic switch between the phases with low- and high
reactivity. This switch is related to the periodic surface transformation between the hex and
($1 \times 1$) surface structures. We note that in contrast to the adsorbed CO which activates the 
hex$\rightarrow$($1 \times 1$) transformation in the considered temperature
interval 400-500~K, the coadsorbed oxygen plays a role of inhibitor for such a 
transformation.
The inhibiting function in this case is based on the fact that the oxygen
predominantly adsorbs in the unreconstructed ($1 \times 1$) phase and reacts with the adsorbed CO. 
The consequent decrease of the CO coverage during the reaction drives the reverse
($1 \times 1$)$\rightarrow$hex surface reconstruction.
In this context, the inhibiting character of the oxygen is a basis for
the oscillatory behavior in the modelling of the surface CO oxidation.

\subsection{Oscillating behavior}

To study the role of the adsorbed oxygen, we introduce the additional equation for the description of the oxygen
coverage $B=\Theta_{\rm O}$ on the Pt surface:

\begin{eqnarray} \label{b}
\frac{\partial B}{\partial t}=p_B k_B s_B\cdot n \left(1-\frac{A}{A_s}-\frac{B}{B_s}\right)^2 -r\cdot AB
\end{eqnarray}
where $p_B=p_{{\rm O}_2}$ and $k_B$ refer to the O$_2$ partial pressure and impingement rate,
respectively. As the O$_2$ sticking coefficient in the hex phase is negligibly small,
the surface state parameter $n$ in the first term restricts the
oxygen adsorption to the unreconstructed ($1 \times 1$) parts of substrate, which occurs 
with the sticking coefficient $s_B=0.3$. 
The mean-field factor $(1-A/A_s-B/B_s)^2$ accounts for the condition that two
neighbouring surface positions not occupied by CO$_{ads}$ and O$_{ads}$ are
required for the dissociative adsorption of O$_2$. 
The second term in (\ref{b}) describes the decrease of B due to the
reaction with CO$_{ads}$, with the reaction rate $r=r_0\exp (-E_r/RT)$. The
similar reaction term is also included into the kinetic equation (\ref{a}) for the
CO coverage. In our analysis of CO oxidation, we choose the values 
$r_0=2\cdot 10^{10}$s$^{-1}$ML$^{-1}$ and $E_r=24.1$~kcal/mol for the reaction parameters.
As the diffusion coefficient of adsorbed oxygen is about three-four orders of magnitude lower
than the CO diffusion parameter $D_A$, the adsorbed oxygen is considered as 
immobile \cite{imbihl3}.

After introducing the dimensionless forms for the oxygen coverage $b=B/B_s$, 
oxygen partial pressure $\tilde{p}_B=p_B k_B s_B/\gamma_{01}B_s$ and
for the reaction rate $\tilde{r}=r/\gamma_{01}$, the equation (\ref{b}) can be
rewritten in the form
\begin{eqnarray} \label{bt}
\frac{\partial b}{\partial \tilde{t}}=f_3(a,n,b)=\tilde{p}_B \cdot n (1-a-b)^2 -\tilde{r}
A_s a\cdot b.
\end{eqnarray}
In the equation (\ref{an}), the modified function $f_1(a,n,b)$ now includes the reaction term: 
$f_1=\tilde{p}_A (1-a) s_A -d_A a -\tilde{r} B_s a\cdot b$.

To analyze the stability of the macroscopically homogeneous solutions of (\ref{an}) and (\ref{bt}), 
we performed a linearization of the system in the vicinity of
the stationary states $\ve{\xi}_0=(a_0, n_0,
b_0)^T$. In this approach, the small deviations $\delta \ve{\xi}=(\delta a,
\delta n, \delta b)^T$ from $\ve{\xi}_0$ can be determined from the system of equations
\begin{eqnarray} \label{stab_an}
\delta \dot{\ve{\xi}}=G \cdot \delta \ve{\xi},
\end{eqnarray}
were $G=\{g_{i,l}\}$ is the matrix of the derivatives 
$g_{i,l}=\partial f_i/\partial \xi_l|_{\ve{\xi}_0} $.

Furthermore, the stability analysis can be reduced to the calculation of the
Lyapunov exponents $\lambda_j$ (eigenvalues of $G$) with the corresponding
deviation vectors (eigenvectors) $\delta \ve{\xi}_j$, so that $G\cdot \delta
\ve{\xi}_j=\lambda_j \ve{\xi}_j$ ($j=1,2,3$). From
the equations (\ref{stab_an}), we can easily obtain the expressions for the 
time evolution of the
eigenvectors $\delta \ve{\xi}_j$: $\delta \ve{\xi}_j=\ve{c}_j^0 \exp(\lambda_j
t)$ where the constants $\ve{c}_j^0$ should be found from the initial
conditions. As a consequence, each arbitrary deviation $\delta \ve{\xi}(t)$ can be
represented as $\delta \ve{\xi}=\sum_j c_j \delta \ve{\xi}_j$ and in this way 
is fully determined by $\lambda_j$.

\begin{figure}[ht]
\epsfxsize=8.0cm {\epsffile{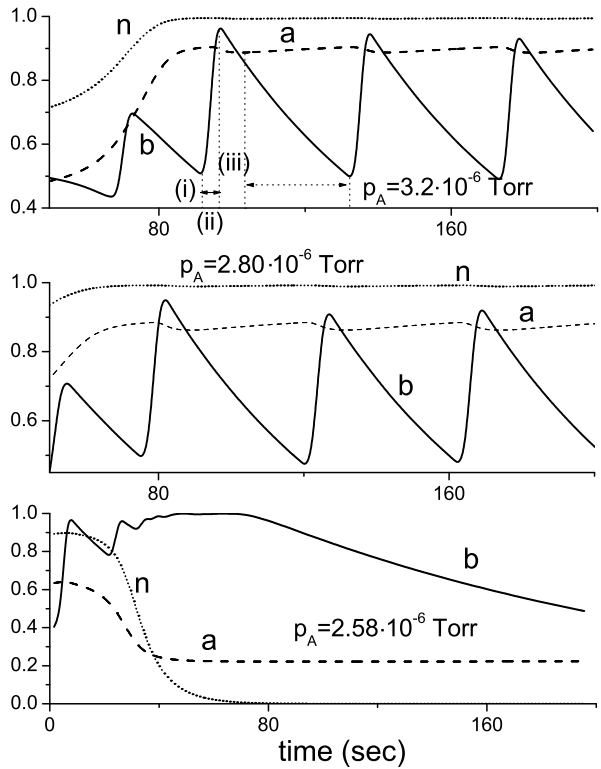}} \caption{Time evolution of dimensionless CO and oxygen
coverages $a$ and $b$ and of the surface state parameter $n$ at various CO pressures
$p_{\rm CO}=p_A$. Here $p_{{\rm O}_2}=p_B=5\cdot 10^{-5}$~Torr, $k_B=5.6\cdot
10^5$ML$\cdot$Torr$^{-1}$s$^{-1}$, $\gamma=0.5$ and $T=450$~K. 
The stages (i)--(iii) on the top panel form a single oscillation
cycle.}\label{fig6}
\end{figure}

In the analysis of the monostable, bistable and oscillation states, the
Lyapunov exponents have been calculated numerically. Fig.~\ref{fig6} shows an
example of the oscillating behavior in the vicinity of a subcritical Hopf
bifurcation point. Here, for larger $p_A$, the real part of two
complex conjugated Lyapunov exponents ${\rm Re} \lambda_2={\rm Re} \lambda_3 >
0$ corresponds to an unstable limit cycle in the phase space determined by the 
vectors $\ve{\xi}=(a,n,b)^T$. 
Such a cycle involves several kinetical stages which are based on a complex interplay
between the CO activator-induced hex$\rightarrow$($1 \times 1$) transformation, 
and the inhibitor-caused reverse decrease of
the adsorbate coverages resulting in the hex surface reconstruction.

Fig.~\ref{fig6} demonstrates how the periodic oscillations of the
coverages $a$ and $b$ are related to the surface transformation described by
the parameter $n$. The increase of the oxygen coverage $b$ up to a saturation
value $b=1$ results in a suppression of the oxygen adsorption (stage (i) in the top panel in
Fig.~\ref{fig6}). Consequently, the reaction step 
becomes dominating and leads to
a decrease of the oxygen coverage $b$. During the CO oxidation, the decreasing 
CO coverage $a$ results in the reverse hex reconstruction of surface, 
a process reflected by a slight decrease of
the state parameter $n$ (stage (ii) in Fig.~\ref{fig6}). 
The reconstruction continues until the
CO adsorption begins to prevail and the CO-activated
hex$\rightarrow$($1 \times 1$)-transformation starts again (stage (iii) in Fig.~\ref{fig6}).
At this stage, the increase of the oxygen adsorption in the ($1 \times 1$) phase
results in a repeat of the oscillation cycle.

As is demonstrated in the bottom panel of Fig.~\ref{fig6}, 
the decrease of the CO pressure $p_A$ leads to the disappearance of the oscillations
and to a further convergence of the system state to a monostable hex phase.

\begin{figure}[ht]
\epsfxsize=8.0cm {\epsffile{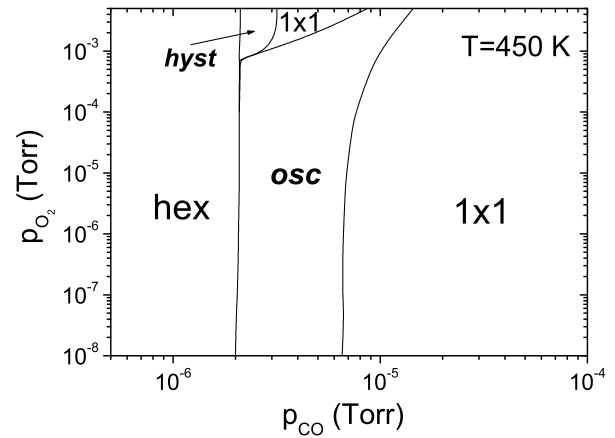}} \caption{Bifurcation phase diagram
($p_{\rm CO}$, $p_{{\rm O}_2}$) calculated at $T=450$.}\label{fig7}
\end{figure}

Fig.~\ref{fig7} presents a bifurcation phase diagram ($p_A$, $p_B$) calculated
at $T=450$~K with the indicated areas of monostable, bistable and oscillating
behavior. One can see that the oscillating region extends with the increasing
$p_B$ which supports a key role of the oxygen in the development of the
oscillations. Another remarkable feature is that the region where the
oscillations occur coincides with the region of the structural surface
transformation which is in full agreement with the experimental observations by
LEED \cite{eiswirth}. 

On the diagram shown in Fig.~\ref{fig7}, 
the oscillating regime disappears in the limit $p_B \rightarrow 0$ which corresponds
to the absence of the adsorbed oxygen on the surface. 
It should be noted that even at extremely low oxygen partial
pressures $p_B < 10^{-7}$~Torr, the calculations still give a region
with low oscillation amplitudes induced by small nonzero $b$. This
is in contradiction to the experimental indications of the disappearance
of oscillations for $p_B < p_A$, a property explained
in terms of a blocking of the adsorption sites by CO \cite{eiswirth}.
Due to a mean-field form of the equations used in our modelling, even at low pressures
$p_B<10^{-7}$~Torr we find a small nonzero $b$ which leads to
the existence of the oscillations. To improve the results in this range 
of pressures, more precise methods have to be applied. A possible way to 
account for the blocking of the adsorption sites by CO
would be to perform Monte Carlo simulations 
which will require further numerical investigations.

The crucial difference between the oscillations on Pt(100) and Pt(110) is related to their
spatio-temporal behavior. While Pt(110) oscillations arise due to the coupling through 
the gas phase and have a well developed regular character, the oscillations on Pt(100) are in
general irregular. This irregularity is usually explained by the inhomogeneous nature of
Pt(100) surfaces, where the waves of structural transformation develop on structurally different
surface patches at different time intervals \cite{eiswirth}. Due to a weak
spatial coupling via the surface diffusion, the
resulting integrated oscillation profiles are nonuniform and highly irregular\cite{imbihl2}.

To gain more insight into the nature of irregularities on Pt(100), one needs to study
comprehensively the oscillations on surfaces which
contain patches of different geometries and characterized by different levels of 
adsorbate coverages. In fact, such studies can be considered in a more wide context, since the modern
lithographic techniques give a possibility to prepare
prepatterned surfaces\cite{imbihl}. Moreover, the control of
adsorbate coverages can be achieved by the methods of scanning tunneling microscopy 
and laser induced thermal desorption which
allow to create artificial micrometer-size microstructures on the surfaces \cite{barth}. 
In this way, the character of the oscillating behavior can be probed and externally tuned.

\section{CONSEQUENCE OF SURFACE INHOMOGENEITIES: CHEMICAL TURBULENCE}

To study the
effect of inhomogeneities, we consider a one-dimensional substrate of a size $L_x$.
The substrate contains a partially reconstructed central surface region $\mathcal{C}$ 
surrounded by the ($1 \times 1$)-surface parts where the hex phase is fully lifted.
For $x \in \mathcal{C}$, the initial surface state is described as
$n(x,t=0)=n_p < 1$. On the substrate, the region $\mathcal{C}$ has the 
length $2\Delta \le L_x$ 
and is defined as $|x-x_c|\le \Delta$ where $x_c$ is the center of the substrate.
In our analysis, we choose the following initial conditions for the surface
coverages:

\begin{eqnarray} \label{ab_inh}
b(x,t=0)=\left\{\begin{array}{cc} b_p, & x \in \mathcal{C}\\
                         0, & |x-x_c| > \Delta \end{array} \right.; \, a(x,t=0)=a_p
\end{eqnarray}

\begin{figure}[ht]
\epsfxsize=8.0cm {\epsffile{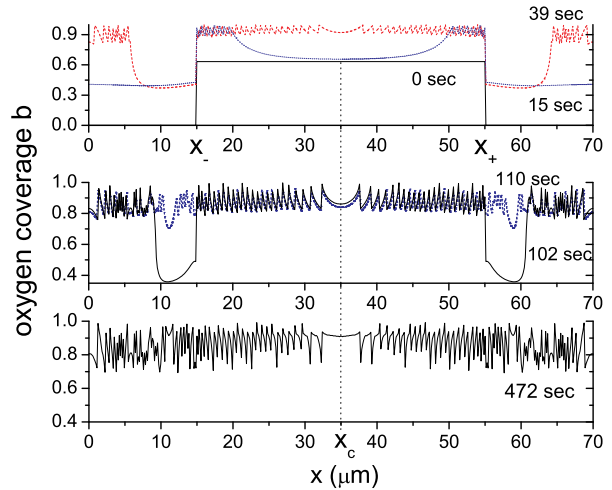}} \caption{Time evolution of spatial
profiles of the oxygen coverage $b$. Here the initial distribution of $b$ is
step-like with $b=b_p=0.63$ inside the central region  $\mathcal{C}$ of a length
$2\Delta=40\mu$m with $x_c=35$~$\mu$m and $b=0$ beyond this region. 
Here $p_A=6.5 \cdot 10^{-6}$~Torr, $p_B=10^{-5}$~Torr, $\eta=0.005$, and 
$T=460$~K.}
\label{fig8}
\end{figure}

\begin{figure}[ht]
\epsfxsize=8.0cm {\epsffile{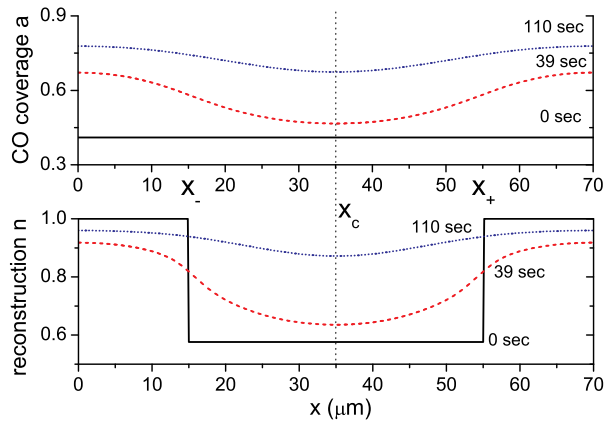}} \caption{Time evolution of spatial
profiles of the surface geometry described by $n$ and of the CO coverage $a$.
The initial $n$ distribution is step-like with $n=n_p=0.57$ inside the central
region $\mathcal{C}$ of a length $2\Delta=40\mu$m with $x_c=35$~$\mu$~m and
$n=1$ beyond this region. The initial CO coverage $a(x)=a_p=0.41$ is
homogeneous. Here $p_A=6.5 \cdot 10^{-6}$~Torr, $p_B=10^{-5}$~Torr, $\eta=0.005$, 
and $T=460$~K.} \label{fig9}
\end{figure}

\begin{figure}[ht]
\epsfxsize=8.0cm \epsffile{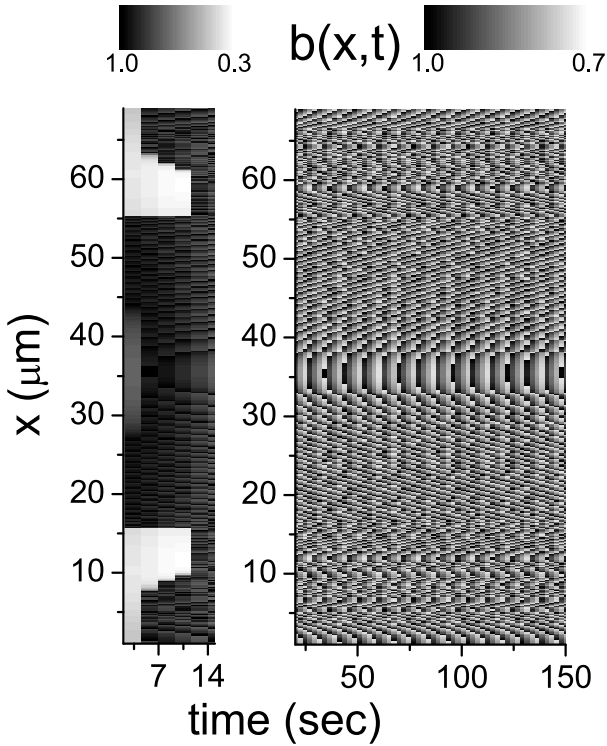} \caption{Amplitude map of oxygen
coverage $b$ in the ($t$, $x$) plane in the
regime of inhomogeneous oscillations. Here $p_A=6.5 \cdot 10^{-6}$~Torr, 
$p_B=10^{-5}$~Torr, and $T=460$~K. The left panel shows the development of the
central nonuniform pattern and the expansion of irregular oscillations 
near the substrate boundaries at the initial stages of temporal evolution.}\label{fig10}
\end{figure}

Inside $\mathcal{C}$, the temperature, reaction/diffusion parameters and the initial conditions
$\ve{\xi}_p=(a_p,n_p,b_p)^T$ correspond to the homogeneous oscillating state
close to the subcritical Hopf bifurcation point.
This implies that in the case when
$\mathcal{C}$ will cover the entire substrate ($2\Delta=L_x$), the time behavior
is characterized by the homogeneous periodic oscillations of the state parameter
$n(x,t)=n(t)$ and of the surface coverages $a(x,t)=a(t)$ and $b(x,t)=b(t)$ already
considered in previous section. In contrast, in the case when the
oscillating surface region $\mathcal{C}$ is located inside the unreconstructed
surface area, the gradients of the adsorbate coverages and of the state parameter $n$ 
near the hex/($1 \times 1$)-interfaces at $x=x_{\pm}$ 
lead to the transition to a highly nonuniform state. 
To see this, in Fig.~\ref{fig8} and Fig.~\ref{fig9} we present the
evolution of the substrate geometry and adsorbate coverages
for a case when $2\Delta=40~\mu$m$<L_x=70\mu$m. One can clearly observe 
the development of
completely new inhomogeneous surface state. In this state, the initial spatial
plateau-like distribution of $b$ is broken. Moreover, the profiles of the oxygen coverage $b$
can be identified as inhomogeneous oscillating standing waves. At the first stages
of the time evolution ($t < 38$~sec in Fig.~\ref{fig8}), these standing patterns are localized
predominantly within $\mathcal{C}$. Furthermore, the additional
irregular patterns develop near the substrate boundaries due to strongly 
increased $b$, which leads to a final extension of  
the inhomogeneous oscillations well beyond $\mathcal{C}$ on the entire substrate
(cases $t=110$ and $472$~sec in Fig.~\ref{fig8}). 
The breaking of the
homogeneous oscillating regime corresponds to the Benjamin-Feir instability of 
homogeneous ($k \rightarrow 0$) oscillating waves. The appearing
irregular spatial patterns are characterized by a deformation of the
wave fronts observed in the amplitude maps of $b$ in Fig.~\ref{fig10}. These 
deformations are related to a transition to the regime of phase turbulence~\cite{kuramoto}.

To deeper understand the origin of the standing waves, let us consider the
equations (\ref{an}), (\ref{bt}) governing the spatio-temporal evolution of the
system. In the development of the wave
instability, the property of the CO coverage $a$ to adjust the
surface state plays a key role. This is demonstrated by the temporal evolution of
$a$ and $n$
presented in Fig.~\ref{fig9}. One can clearly see the development of inhomogeneous 
spatial profiles from the initial uniformly distributed $a(x)=a_p$. 
The decrease of $a$ inside $\mathcal{C}$ is produced by the
initial plateau-like distribution of $n$ in this region. As a consequence, the
parabolic concave shape of $a(x)$ leads to a negative second
derivative $\partial^2 a/\partial x^2$ close to
$x=x_{\pm}$. To see the effect of $\partial^2 a/\partial x^2< 0$ on the wave properties, we
consider a modified equation for $b$ which can be derived in the vicinity of $\ve{\xi}_p$
on the basis of (\ref{an}) and (\ref{bt}):

\begin{eqnarray} \label{bt2}
\frac{\partial^2 b}{\partial t^2}\approx \frac{\partial^2 a}
{\partial x^2} (2n_p \tilde{p}_B-\tilde{r}A_s) b+f_0.
\end{eqnarray}
Here $f_0 \approx -2n_p \tilde{p}_B (1-a) \partial^2 a/\partial x^2$.
In the region $\mathcal{C}$, the expression for ${\partial^2 a}/{\partial x^2}$ 
can be approximated by a parabola: 
\begin{eqnarray} \label{d2a}
\partial^2 a/\partial x^2 \approx -\alpha_1(x-x_c)^2+\alpha_0,
\end{eqnarray}
with the curvature determined by the parameter $\alpha_1$.

Furthermore, close to the hex/$1 \times 1$-interfaces, 
we have $-\alpha_1 (x_c-x)^2+\alpha_0=-\alpha(x)
< 0$. As a result, in the interface region the time-dependent part of the
solution of (\ref{bt2}) has the form of a standing wave
\begin{eqnarray} \label{bx}
&& b(x,t) = b_0 \cos(\eta(x)t+\eta_0),\\
&& \eta(x)=\sqrt{\alpha(x) (2n_p \tilde{p}_B-\tilde{r}A_s)}. \nonumber
\end{eqnarray}
The parameters $\eta_0$ and $b_0$ can be found from the boundary conditions
$b(x_{\pm},t)=b_{>}\cos(\omega t)$. Here $b_{>}(t)=b_{>}\cos(\omega t)$ is the
oxygen coverage in the ($1\times 1$) areas beyond $\mathcal{C}$ which is assumed to be
homogeneous for simplicity. The frequency $\omega$ is the oscillation frequency
of the system. With these boundary conditions, close to the
interface ($|x-x_c| \sim \Delta$) we have
\begin{eqnarray} \label{bx2}
&& b(x,t) = b_{>} \cos(\omega t+\delta \phi(x,t)),
\end{eqnarray}
where the function $\delta \phi(x,t)\sim \alpha_1 (\Delta^2-(x-x_c)^2)t/\Delta$
describes the deformation of the homogeneous wave profile due to a nonzero
curvature $\alpha_1$ from (\ref{d2a}). 
As the phase deformation $\delta \phi(x,t)$ is controlled by
$\alpha_1$, the oscillations of $\alpha_1(t)$ during the periodic cycles 
result in the inhomogeneous wave
front oscillating with the constant frequency $\omega$ which is 
shown in the right panel of Fig.~\ref{fig10}.

In order to test the
Benjamin-Fair instability characterized by unstable oscillating state with 
respect to inhomogeneous perturbations, we have calculated the Floquet
exponents for different wave numbers $k$. For the analysis of the spatially inhomogeneous
states, we introduce the Fourier transforms $\xi_k=(a_k, n_k, b_k)^T$ and $f_i(k)$
\begin{eqnarray} \label{fk}
f_i(k)=\frac{1}{N_x} \sum_x f_i(a,n,b) \exp(-ikx)=f_i(a_k,n_k,b_k)
\end{eqnarray}
where the range of the wave numbers $k=2\pi l/L_x$ ($l=1,\ldots,N_x$) is determined by the periodic
boundary conditions $\ve{\xi}(x)=\ve{\xi}(x+L_x)$.

For a state represented by the vector $\ve{\xi}^0=(a_0,n_0,b_0)^T$ oscillating with a
period $T$ and described by (\ref{an}) and (\ref{bt}), the analysis of the Floquet
exponents can be reduced to the solution of the eigenvalue problem with the
Jacobian $G_k=\{ g_{i,j}(k)\}$ where $g_{i,j}(k)=\partial f_i(k)/\partial
\xi_{kj}|_{\ve{\xi}_k^0}$. The condition $\lambda_k < 0$ 
implies the stability of $\ve{\xi}^0$ with respect to small inhomogeneous
perturbations $\delta\ve{\xi}_k$. In contrast to this, the opposite condition
$\lambda_k > 0$ corresponds to the instability of the
homogeneous oscillating state and to the appearance of modulation in
the system.

\begin{figure}[ht]
\epsfxsize=8.0cm {\epsffile{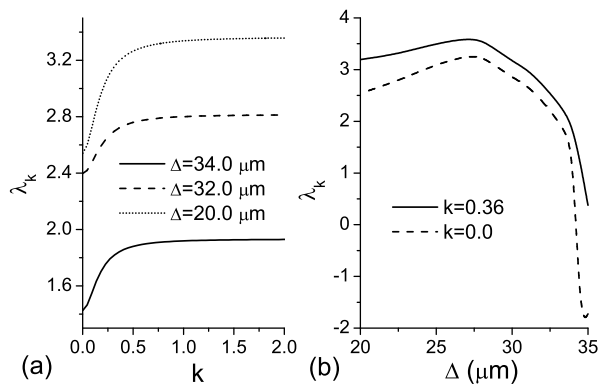}} \caption{(a) Dispersion of the real part of the Floquet 
exponent $\lambda_k$ for different values of the half-width $\Delta$ of the reconstructed region
$\mathcal{C}$. (b) $\lambda_k$ as a function of $\Delta$ for different $k$. 
Here $p_A=6.4 \cdot 10^{-6}$~Torr, $p_B=10^{-5}$~Torr, and $T=460$~K.} 
\label{fig11}
\end{figure}

Fig.~\ref{fig11} presents the behavior of the maximal real part of the Floquet exponent
$\lambda_k=\max\{{\rm Re}\lambda_k^1,{\rm Re}\lambda_k^2,{\rm Re}\lambda_k^3\}$
which occurs under a variation of the length $\Delta$.
One can clearly see that the decrease of $\Delta$
leads to a strong increase of $\lambda_k$ for all values of $k$
(Fig.~\ref{fig11}(a)). The positive $\lambda_k$ for nonzero $k$
indicates the formation of the inhomogeneous state presented in
Fig.~{\ref{fig8}} and Fig.~{\ref{fig10}}. Moreover, Fig.~\ref{fig11}(b)
shows that the bulk-like oscillation state which develops in an initially homogeneous 
region $\mathcal{C}$ of a maximal length 
$\Delta=35 \mu$m, is unstable with respect to inhomogeneous perturbations.
On such a substrate, the uniform oscillations
cannot be destroyed by homogeneous perturbations which is reflected 
in $\lambda_k(k=0)<0$ in Fig.~\ref{fig11}(b). In contrast, due to the 
positive $\lambda_k (k\ne 0)$,
the inhomogeneous perturbations diverge and lead to spatially modulated oscillatins.

The principal feature of $\lambda_k$ is its almost dispersionless (flat) 
character which can be clearly seen in Fig.~\ref{fig11}(a). 
Specifically, for all values of $\Delta$ we obtain an initial increase of 
$\lambda_k$ at small $k$,
with a further saturation to the maximal constant value which occurs for larger
$k > 0.5$ (Fig.~\ref{fig11}(a)). 
This "flat-band"-character is in strong contrast to the dispersion $\lambda_k$ usually obtained 
in the modelling of the global coupling for Pt(110) \cite{falcke,bar2}. In the latter case, $\lambda_k$
has a distinct well-defined maximum at a certain $k=k_0$ which corresponds to the appearance of spatial 
clusters with $k_0$-modulation. 
In our case, due to the flatness of $\lambda_k$ at larger $k$, no distinctly leading
mode can be formed. Instead, all possible $k$-modulations can exist in such an instable system. 
This is the key reason for
a highly irregular form of the standing waves in Fig.~\ref{fig8}. Although keeping the basic features
related to the central symmetry, the irregular wave profiles are strongly sensitive to the initial state
of substrate and to the initial spatial distributions of adsorbate coverages, and in this sence they are
random. In the real systems, the type of irregular profile can be pinned by surface defects
like steps and surface imperfections, with possible further random changes caused by the
noise in the system.

In fact, the obtained spatio-temporal evolution of the irregular oscillations on the
substrate is consistent with the the development of irregular patterns on spatially
separated patches typically observed on Pt(100).
It should be noted that the inhomogeneous oscillating states and standing waves have been previously
discussed for CO oxidation on Pt(110) where the main mechanism responsible for
these phenomena is the coupling through the gas phase \cite{falcke,bar2,mertens}. In
contrast to Pt(110), on Pt(100) the inhomogeneities in the surface
geometry play a central role. Such inhomogeneities lead to the nucleation of
adsorbate islands and to a decisive role of the
boundaries between geometrically different surface areas 
in the formation of standing waves and irregular oscillations.

\section{CONCLUSIONS}

We have studied the CO adsorption and catalytic CO oxidation on inhomogeneous
Pt(100) surfaces which contain structurally different hex and ($1\times 1$)-domains. 
In our model, the structural surface transformation is described in terms of a bistability
characterized by a coexistence of the reconstructed and $1 \times 1$ surface areas.
The bistable region terminates in the critical cusp point which can be approached at
higher temperatures and pressures. Due to significant
increase of the fluctuations, the behavior in the vicinity of the cusp point should be
qualitatively different from the hysteresis-type behavior 
typically observed in the experiments 
under ultrahigh vacuum conditions.    
We have also studied the nucleation and growth of CO islands on inhomogeneous substrate,
a process driven by the difference in 
the desorption properties of CO in the hex and $1 \times 1$ phases. 
We have also performed the analysis of the oscillation behavior on the surface
during the CO oxidation. The surface inhomogeneities lead to the Benjamin-Feir instability
and to the irregular standing waves of adsorbate coverages.
The obtained deformations of the wave fronts is a direct consequence of structural 
inhomogeneities typical for Pt(100) which allows to gain deeper insight into the 
nature of irregular oscillations on this type of surfaces.


\end{document}